\documentclass[prd,aps,showpacs,nofootinbib,eqsecnum,onecolumn,groupedaddress,
amssymb]{revtex4}
\usepackage{graphicx}
\usepackage[english]{babel}
\usepackage{amsmath}
\usepackage{amssymb}
\usepackage{amsfonts}
\usepackage{color,amsxtra}
\usepackage{epsf}
\usepackage{enumerate}
\usepackage{hhline}
\usepackage{array}
\usepackage{tabularx}
\usepackage{subfigure}
\usepackage{fancyhdr}
\usepackage{mathrsfs}
\newcommand{\be}{\begin{equation}}
\newcommand{\ee}{\end{equation}}
\newcommand{\bea}{\begin{eqnarray}}
\newcommand{\eea}{\end{eqnarray}}
\newcommand{\beaa}{\begin{eqnarray*}}
\newcommand{\eeaa}{\end{eqnarray*}}

\newcommand{\e}{\mathrm{e}}

\newcommand{\Eqn}[1]{&\hspace{-0.2em}#1\hspace{-0.2em}&}

\def\be{\begin{equation}}
\def\ee{\end{equation}}
\def\bea{\begin{eqnarray}}
\def\eea{\end{eqnarray}}

\def\e{\mathrm{e}}

\begin{document}

\title{Inflation in a conformally-invariant two-scalar-field theory with 
an extra $R^2$ term}

\author{Kazuharu Bamba$^{1, 2, 3}$, Sergei D. Odintsov$^{4, 5}$ and 
Petr V. Tretyakov$^{6}$
}
\affiliation{
$^1$Division of Human Support System, Faculty of Symbiotic Systems Science, Fukushima University, Fukushima 960-1296, Japan\\
$^2$Leading Graduate School Promotion Center,
Ochanomizu University, Tokyo 112-8610, Japan\\
$^3$Department of Physics, Graduate School of Humanities and Sciences, Ochanomizu University, Tokyo 112-8610, Japan\\ 
$^4$Institut de Ciencies de lEspai (IEEC-CSIC), 
Campus UAB, Carrer de Can Magrans, s/n 
08193 Cerdanyola del Valles, Barcelona, Spain\\
$^5$Instituci\'{o} Catalana de Recerca i Estudis Avan\c{c}ats
(ICREA), Passeig Llu\'{i}s Companys, 23 08010 Barcelona, Spain\\ 
$^6$Joined Institute for Nuclear Research, Dubna, Moscow Region, Russia
}


\begin{abstract} 
We explore inflationary cosmology in a theory where there are 
two scalar fields which non-minimally couple to the Ricci scalar 
and an additional $R^2$ term, which breaks the conformal invariance. 
Particularly, we investigate the slow-roll inflation 
in the case of one dynamical scalar field and 
that of two dynamical scalar fields. 
It is explicitly demonstrated that the spectral index of scalar mode of 
the density perturbations 
and the tensor-to-scalar ratio can be consistent with 
the observations acquired by the recent Planck satellite. 
The graceful exit from the inflationary stage is achieved as in convenient 
$R^2$ gravity. 
We also propose the generalization of the model under discussion with 
three scalar fields. 
\end{abstract}

\pacs{98.80.-k, 98.80.Cq, 04.50.Kd, 12.60.-i}
\hspace{12.5cm} FU-PCG-03, OCHA-PP-334

\maketitle

\def\thesection{\Roman{section}}
\def\theequation{\Roman{section}.\arabic{equation}}

\section{Introduction}

The natures on inflation~\cite{Inflation, N-I, Starobinsky:1980te} in the early universe have been revealed by the recent cosmological observations 
such as the Wilkinson Microwave anisotropy probe (WMAP)~\cite{Komatsu:2010fb, Hinshaw:2012aka}, 
the Planck satellite~\cite{Planck:2015xua, Ade:2015lrj}, 
and the BICEP2 experiment~\cite{Ade:2014xna, Ade:2015tva}
on the quite tiny anisotropy of the cosmic microwave background (CMB) radiation. Owing to the release of the recent observational data, 
in addition to seminal inflation with single scalar field 
such as new inflation~\cite{N-I}, chaotic inflation~\cite{Linde:1983gd}, 
natural inflation~\cite{Freese:1990rb}, and power-law inflation with the exponential inflaton potential~\cite{Yokoyama:1987an}, 
novel models of single field inflation 
have been proposed in Refs.~\cite{R-I-S, Bezrukov:2007ep}\footnote{Recently, there have also been studied inflationary models with two/multiple scalar fields (or a complex scalar field with two scalar degrees of freedom) 
such as hybrid inflation~\cite{Linde:1993cn} and a kind of its extensions~\cite{H-I}.} (for reviews on more various inflationary models, see, e.g.,~\cite{Lidsey:1995np, Lyth:1998xn, Gorbunov:2011zzc, MRV, Linde:2014nna}). 

In addition to inflationary models driven by the scalar filed (i.e., 
the inflaton field) described above, there have been considered 
the so-called Starobinsky inflation~\cite{Starobinsky:1980te, Vilenkin:1985md} 
originating from the higher-order curvature term such as $R^2$ term\footnote{Note that the Starobinsky or $R^2$ inflation in the case of no matter is equivalent to non-minimal Higgs inflation considered in Ref.~\cite{Bezrukov:2007ep}.}, 
where $R$ is the Ricci scalar. 
This model is observationally supported by the Planck results. 
Such a theory can be interpreted as a kind of modified gravity theories 
including $F(R)$ gravity to account for the late-time cosmic acceleration 
(for reviews on dark energy problem and modified gravity theories, 
see, for instance,~\cite{Nojiri:2010wj, Nojiri:2006ri, Joyce:2014kja, Book-Capozziello-Faraoni, Capozziello:2011et, Koyama:2015vza, Bamba:2012cp, delaCruzDombriz:2012xy, Bamba:2015uma, Bamba:2013iga, Bamba:2014eea}). 
Various inflationary models in modified gravity theories corresponding 
to extensions of the Starobinsky inflation have been explored 
in Refs.~\cite{Inflation-M-G, LT}. 

In this paper, we investigate inflation in a theory consisting of 
two scalar fields which non-minimally couple to the Ricci scalar 
and an additional $R^2$ term\footnote{In Ref.~\cite{Bamba:2006mh}, 
inflationary cosmology has been studied in 
a theory with two scalar fields non-minimally coupling 
to the Ricci scalar.}. 
We consider the conformally-invariant two-scalar-field theory 
in which the conformal invariance is broken by adding an $R^2$ term. 
In particular, we explore the slow-roll inflation 
in the cases of (i) one dynamical scalar field (namely, we set 
one of two scalar fields a constant) and (ii) two dynamical 
scalar fields. 
As a consequence, we analyze  
the spectral index of scalar mode of the density perturbations 
and the tensor-to-scalar ratio and 
compare the theoretical results with 
the observational data obtained by the recent Planck satellite 
and the BICEP2 experiment. 
It is clearly shown that the spectral index 
and tensor-to-scalar ratio can be compatible with 
the recent Planck results. 

The motivation to propose our theory is to unify inflation in the early universe originating from the $R^2$ term and the late-time cosmic acceleration, i.e., the dark energy dominated stage with dark matter. The $R^2$ term is interpreted as the contribution from modified gravity, dark energy is described by 
one of the scalar fields, and dark matter is represented by 
the other scalar field. 
Furthermore, it seems that multiple field inflation models can fit the Planck data better than single field inflation models. 
Inflationary models with multi-scalar fields~\cite{M-F-I} 
including the so-called curvaton scenario~\cite{C-S}
have been constructed and the cosmological perturbations 
in these models have also been investigated~\cite{CP-MS}. 
We use units of $k_\mathrm{B} = c = \hbar = 1$ and express the
gravitational constant $8 \pi G_\mathrm{N}$ by
${\kappa}^2 \equiv 8\pi/{M_{\mathrm{Pl}}}^2$ 
with the Planck mass of $M_{\mathrm{Pl}} = G_\mathrm{N}^{-1/2} = 1.2 \times 
10^{19}$\,\,GeV. 

The organization of the paper is the following. 
In Sec.~II, we explain our model action and derive the gravitational field equation and the equations of motions for the scalar fields. 
We also examine the slow-roll inflation in the case of one dynamical 
scalar field and study the dynamics of the system including 
the equilibrium points in detail. 
In Sec.~III, with the conformal transformation~\cite{M-FM}, we explore 
inflationary cosmology in the Einstein frame. 
Especially, we study inflationary models in the case that the conformal 
scalar is the dynamical inflaton field and other scalar fields are set 
to be constants. 
In Sec.~IV, we investigate the slow-roll inflation in the case 
of two dynamical scalar fields. Particularly, we consider 
the resultant spectral index and the tensor-to-scalar ratio 
in the Jordan frame (i.e., the original conformal frame). 
In Sec.~V, we explore the graceful exit from inflation, namely, 
the instability of the de Sitter solution in the present theory. 
As a demonstration, we concentrate on the case of one dynamical scalar field 
in the Einstein frame, because, as is described in Sec.~III, 
in this case the spectral index and the tensor-to-scalar ratio 
can be consistent with the recent Planck results. 
Finally, conclusions are described in Sec.~VI. 

\section{Two scalar field model with breaking the conformal invariance} 

\subsection{Model action and its transformation into the canonical form} 

Our model action, which consists of two scalar fields $\phi$ and $u$ and an 
$R^2$ term, is described as
\begin{equation}
S= \int d^4 x \sqrt{-g}\left\{ \frac{\alpha}{2} R^2 + \frac{s}{2}\left[ \frac{(\phi^2-u^2)}{6}R + (\nabla\phi)^2 - (\nabla u)^2 \right] - (\phi^2-u^2)^2J(y) \right\}\,,
\label{1.1}
\end{equation}
where $g$ is the determinant of the metric $g_{\mu\nu}$, $R$ is the scalar curvature, $\alpha (\neq 0)$ is a non-zero constant, $s=\pm 1$ 
is a model parameter, 
$\nabla$ is the covariant derivative, and $J (y)$ is a function of $y$ defined 
as $y \equiv u/\phi$. 
The action in Eq.~(\ref{1.1}) without the $R^2$ term has been proposed and studied in Refs.~\cite{Bars:2011th, Bars:2011aa, B-BST-BST, Carrasco:2013hua, Bars:2008sz, Bars:2010zh, Nishino:2004kb, Bamba:2014kza}. 
Here and the following, we have taken $2\kappa^2 = 1$. 
Regarding the property of the action in Eq.~(\ref{1.1}), 
we remark that if there does not exist the $R^2$ term, 
this action is conformally invariant, 
while that when the $R^2$ term is added, 
the conformal invariance of this action is effectively broken. 
We also note that 
our action may be considered to be invariant yet under the 
restricted conformal invariance~\cite{Edery:2014nha}. 
By introducing an auxiliary field $\Phi$, the action in Eq.~(\ref{1.1}) can be rewritten as
\begin{equation}
S= \int d^4 x \sqrt{-g}\left\{ \left[ \Phi + \frac{s}{12}(\phi^2-u^2) \right]R -\frac{\Phi^2}{2\alpha} + \frac{s}{2} \left[  (\nabla\phi)^2 - (\nabla u)^2 \right] - (\phi^2-u^2)^2J(y) \right\}\,.
\label{1.2}
\end{equation}
The simplest way to transform the action in Eq.~(\ref{1.2}) into 
the canonical form is to take the following gauge: 
\begin{equation}
\Phi + \frac{s}{12}(\phi^2-u^2) = 1\,.
\label{1.3}
\end{equation}
With this gauge, the action in Eq.~(\ref{1.2}) is transformed into 
\begin{equation}
S= \int d^4 x \sqrt{-g}\left\{ R -\frac{1}{2\alpha}\left[ 1 - \frac{s}{12}(\phi^2-u^2) \right]^2 + \frac{s}{2}\left[  (\nabla\phi)^2 - (\nabla u)^2 \right] - (\phi^2-u^2)^2J(y) \right\}.
\label{1.4}
\end{equation}

Here, we explicitly explain the properties of the actions described above. 
The representative feature of the action in Eq.~(\ref{1.1}) is that 
this action is not conformally invariant, while it is scale invariant. 
To eliminate the non-minimal coupling between the scalar fields and 
the scalar curvature in the action in Eq.~(\ref{1.2}), we have used the gauge in Eq.~(\ref{1.3}). 
As a result, for the resultant action in Eq.~(\ref{1.4}), 
the scale invariance is broken, although the canonical form, 
i.e., the Einstein-Hilbert term, is recovered. 
Therefore, it is considered that the form of the gauge in Eq.~(\ref{1.2}) has two aspects. One is to make the action canonical, but 
the other is to apparently break its scale invariance of the action. 

{}From the action in Eq.~(\ref{1.4}), we obtain the gravitational equation: 
\begin{eqnarray}
&&
R_{\mu\nu}-\frac{1}{2}Rg_{\mu\nu}+\frac{s}{2}(\nabla_{\mu}\phi\nabla_{\nu}\phi-\nabla_{\mu} u \nabla_{\nu} u) 
\nonumber \\ 
&&
{}+ \frac{1}{2}g_{\mu\nu} \left \{
\frac{1}{2\alpha}\left [1 - \frac{s}{12}(\phi^2-u^2) \right ]^2 - \frac{s}{2}\left[ (\nabla\phi)^2-(\nabla u)^2\right ] +(\phi^2-u^2)^2 J(y)
\right \}=0 \,,
\label{1.7}
\end{eqnarray}
where $R_{\mu\nu}$ is the Ricci tensor, 
and the equations of motion for the scalar fields $\phi$ and $u$:
\begin{eqnarray}
&&
s\Box\phi - \frac{s}{6\alpha}\phi\left [1 - \frac{s}{12}(\phi^2-u^2) \right ] + 4\phi(\phi^2-u^2)J(y) - \frac{u}{\phi^2}(\phi^2-u^2)^2 J'(y)=0 \,,
\label{1.5} \\ 
&&
s\Box u - \frac{s}{6\alpha}u\left [1 - \frac{s}{12}(\phi^2-u^2) \right ] + 4u(\phi^2-u^2)J(y) - \frac{1}{\phi}(\phi^2-u^2)^2 J'(y)=0 \,,
\label{1.6}
\end{eqnarray}
where $\Box \equiv g^{\mu \nu} {\nabla}_{\mu} {\nabla}_{\nu}$
is the covariant d'Alembertian operator for scalar quantities, 
and the prime denotes the derivative with respect to $y$. 
Furthermore, the action in Eq.~(\ref{1.4}) is represented as
\begin{equation}
S= \int d^4 x \sqrt{-g}\left\{ R  + \frac{s}{2} \left[ (\nabla\phi)^2 - (\nabla u)^2 \right] - V(\phi, u, J) \right\}\,. 
\label{1.9}
\end{equation}
Here, $V(\phi, u, J)$ is a potential for $u$ and $\phi$, defined as
\begin{equation}
V(\phi, u, J) \equiv \frac{1}{2\alpha}\left[ 1 - \frac{s}{12}(\phi^2-u^2) \right]^2+(\phi^2-u^2)^2 J(y) \,.
\label{1.10}
\end{equation}
In this case, the gravitational field equation can be expressed by 
\begin{equation}
R_{\mu\nu}-\frac{1}{2}Rg_{\mu\nu}+\frac{s}{2}\left(\nabla_{\mu}\phi\nabla_{\nu}\phi-\nabla_{\mu} u \nabla_{\nu} u\right) + \frac{1}{2}g_{\mu\nu} \left\{ V - \frac{s}{2}\left[(\nabla\phi)^2-(\nabla u)^2\right]
\right\} =0 \,. 
\label{1.13}
\end{equation}
The equations of motion for $u$ and $\phi$ also become quite simple as 
\begin{eqnarray}
s\Box\phi + V_{\phi} \Eqn{=} 0\,,
\label{1.11} \\  
s\Box u - V_{u} \Eqn{=} 0\,,
\label{1.12}
\end{eqnarray}
with $V_{\phi} \equiv \partial V/\partial \phi$ and 
$V_{u} \equiv \partial V/\partial u$. 

The flat Friedmann-Lema\^{i}tre-Robertson-Walker (FLRW) metric 
$ds^2 = -dt^2 + a^2(t) \sum_{i=1,2,3}\left(dx^i\right)^2$ 
is assumed, where $a(t)$ is the scale factor. 
The Hubble parameter is given by 
$H \equiv \dot{a}/a$, where the dot shows the time derivative.  
In this background space-time, the gravitational field equations read 
\begin{eqnarray}
&&
3H^2 + \frac{s}{4}(\dot\phi^2-\dot u^2)-\frac{1}{2}V=0 \,,
\label{1.14} \\ 
&&
2\dot H + 3H^2 - \frac{s}{4}(\dot\phi^2-\dot u^2)-\frac{1}{2}V=0 \,. 
\label{1.15}
\end{eqnarray}
In addition, the equation of motions for $\phi$ and $u$ become 
\begin{eqnarray} 
s(\ddot\phi+3H\dot\phi) - V_{\phi} \Eqn{=} 0\,,
\label{eq:IIA-2.15} \\ 
s(\ddot{u}+3H\dot{u}) + V_{u} \Eqn{=} 0\,.
\label{eq:IIA-2.16}
\end{eqnarray}
%

\subsection{Single dynamical scalar field model}

We consider that $\phi$ is the inflaton field and 
the slow-roll inflation occurs. 
Namely, $\left|\ddot{\phi}\right| \ll \left| 3H \dot{\phi} \right|$ 
in the equation of motion for $\phi$ (\ref{eq:IIA-2.15}) 
and the kinetic energy $\left(1/2\right) \dot{\phi}^2$ of $\phi$ is 
much smaller than the potential energy $V$. 
We also suppose that the mass of the other scalar field $u$ is much smaller than the Hubble parameter at the inflationary stage and hence the amplitude of $u$ does not vary during inflation. 
Accordingly, we set $u=u_0 (= \mathrm{constant})$  
at the inflationary stage. 
In this case, it follows from Eq.~(\ref{1.12}) that 
\begin{equation}
V_{u}=\frac{s}{6\alpha}u\left [1 - \frac{s}{12}(\phi^2-u^2) \right ]- 4u(\phi^2-u^2)J(y) + \frac{1}{\phi}(\phi^2-u^2)^2J'(y) =0\,.
\label{1.16}
\end{equation}
This has to be satisfied for a value of $\phi(t)$ and $u=u_0$, 
although it cannot be true for an arbitrary function $J(y)$. 
In general, we can take $u_0=0$ and $J=J(y^2)$. 
The simplest case is $u_0=0$ and $J=1$.  
For such a case, the scalar field $u$ is totally decomposed 
from equations during inflation.  
Consequently, the effective potential of the inflaton $\phi$ is given by 
\begin{equation}
V_\mathrm{eff}(\phi) = \frac{1}{2\alpha}
\left(1 - \frac{s}{12}\phi^2 \right)^2+C\phi^4 \,, 
\label{1.19}
\end{equation}
with $C$ a constant. This expression is appropriate 
for any function $J=J(y^2)$. 
The number of $e$-folds during inflation 
is represented as 
\begin{equation}
N_e(\phi)=\int^{\phi_\mathrm{f}}_{\phi}H(\hat{\phi})\frac{d\hat{\phi}}{\dot{\hat{\phi}}}
=-\frac{s}{2}\int^{\phi}_{\phi_\mathrm{f}}\frac{V(u, \hat{\phi}, J)}{V_\phi (u, \hat{\phi}, J)}d\hat{\phi} \,,
\label{1.20}
\end{equation}
where $\phi_\mathrm{f}$ is the amplitude of $\phi$ at the end 
of inflation. 
For the effective potential in Eq.~(\ref{1.19}) with 
$\phi \gg \phi_\mathrm{f}$, we have 
\begin{equation}
N_e(\phi)=-\frac{s}{2}\left[\frac{\phi^2}{8} + \frac{432\alpha C \ln(s^2\phi^2+288\alpha C\phi^2 -12s)}{s\left(s^2+288\alpha C\right)} -\frac{3\ln\phi}{s} \right]\,.
\label{1.21}
\end{equation}
Hence, it is seen that there exist the second and third logarithmic correction terms in Eq.~(\ref{1.21}) in comparison with the number of $e$-folds for the quartic inflaton potential as $\lambda\phi^4$ with $\lambda$ a constant. 

The slow-roll parameters in the so-called kinematic approach are defined as (for reviews, see, for instance,~\cite{Lidsey:1995np, Lyth:1998xn})
\begin{eqnarray}
\epsilon \Eqn{\equiv} -\frac{\dot H}{H^2} \,,  
\label{0.1} \\
\eta \Eqn{\equiv} \epsilon - \frac{\ddot H}{2H\dot H} \,.
\label{0.2}
\end{eqnarray}
For the slow-roll inflation, 
the spectral index $n_\mathrm{s}$ of the curvature perturbations and 
the tensor-to-scalar ratio $r$ of the density perturbations 
are described by~\cite{Mukhanov:1981xt, L-L}  
\begin{eqnarray}
n_\mathrm{s} \Eqn{=} 1 -6\epsilon + 2\eta \,,
\label{0.4} \\ 
r \Eqn{=} 16 \epsilon \,. 
\label{0.3} 
\end{eqnarray}
For the slow-roll inflation driven by the effective potential 
in Eq.~(\ref{1.19}), the slow-roll parameters are written as 
\begin{eqnarray}
\epsilon \Eqn{=} -\frac{1}{s}\left( \frac{1}{V_\mathrm{eff}(\phi)} \frac{dV_\mathrm{eff}(\phi)}{d\phi} \right)^2=-\frac{16\phi^2}{s}\left [ \frac{(s^2+288\alpha C)\phi^2-12s}{288\alpha C\phi^4+(12-s\phi^2)^2} \right ]^2 \,,
\label{1.22} \\
\eta \Eqn{=} -\frac{2}{s} \frac{1}{V_\mathrm{eff}(\phi)} 
\frac{d^2 V_\mathrm{eff}(\phi)}{d \phi^2} 
=-\frac{24}{s} 
\frac{(s^2+288\alpha C)\phi^2-4s}{288\alpha C\phi^4+(12-s\phi^2)^2} \,,
\label{1.23}
\end{eqnarray}
where we have used Eqs.~(\ref{1.11}), (\ref{1.14}), and (\ref{1.15}). 
With Eq.~(\ref{1.22}), we find that $s$ has to be negative 
because $\epsilon >0$. 
When $\epsilon <0$, it follows from Eq.~(\ref{0.1}) that $\dot{H} >0$. 
This means that not the slow-roll inflation but the so-called super-inflation can occur. 

If we take the number of $e$-folds, whose value has to be large such as $N_e=50$--$60$ enough to solve the so-called horizon and flatness problem and 
the values of $n_\mathrm{s}$ and $r$ suggested by observations, 
it is apparently regarded that 
there are three equations (\ref{1.21}), (\ref{0.4}) and (\ref{0.3})
for three independent variables ($\phi$, $\alpha$, $C$). 
By solving these three equations, 
we can estimate viable values of our model parameters. 
However, $\alpha$ and $C$ are incorporated into all the equations 
in the form of $x \equiv \alpha C$. 
Hence, it is necessary to analyze a system of two equations, for example, 
Eqs.~(\ref{0.4}) and (\ref{1.21}), whereas $r$ should depend on 
$n_\mathrm{s}$. {}From this reason, we may try to modify the initial action in Eq.~(\ref{1.1}) in the following way. Let us suppose that $s$ is an arbitrary numerical parameter. In this case, we have a system of three equations for three variables. 
We also remark that our potential is well known as a potential of ``Spontaneous symmetry breaking inflation (SSBI)'' and 
a viable inflationary model can be constructed~\cite{MRV}, although only for positive values of the parameter $s$. 

According to the Planck 2015 results, 
$n_{\mathrm{s}} = 0.968 \pm 0.006\, (68\%\,\mathrm{CL})$~\cite{Planck:2015xua, Ade:2015lrj} and $r < 0.11\, (95\%\,\mathrm{CL})$~\cite{Ade:2015lrj}. 
These values are consistent with those obtained by the WMAP 
satellite~\cite{Komatsu:2010fb, Hinshaw:2012aka}. 
The BICEP2 experiment has suggested $r=0.20^{+0.07}_{-0.05}\, 
(68\%\,\mathrm{CL})$~\cite{Ade:2014xna}, 
but recently the $B$-mode polarization of the CMB radiation is considered to 
the contribution from the dust, and not the primordial gravitational 
waves~\cite{Ade:2015tva}. 

\begin{center}
\begin{figure}[tbp]
\includegraphics[width=8cm]{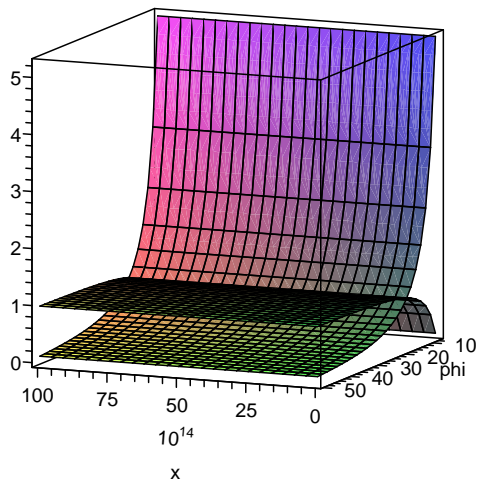} 
\caption{$n_\mathrm{s}$ and $r$ as functions of $x$ and $\phi$ for $s=-1$. 
The sheet whose height changes from $\sim 0$ to $\sim 1$ denotes $n_\mathrm{s}$, while the other sheet whose height varies from $\sim 0$ to $\sim 5$ 
shows $r$.}
\label{nr1}
\end{figure}
\end{center}

In our model, $n_\mathrm{s} \approx 0.96$ can be realized for 
$50 \leq N_e \leq 60$ and a wide range of parameter $s$, but 
$r<0.11$ cannot be satisfied for any values of $s$. 
Therefore, we may put $s=-1$, although 
it is difficult to produce the value of 
$r$ compatible with the Planck/WMAP data in our model. 
In Fig.~\ref{nr1}, we show the values of $n_\mathrm{s}$ and $r$ 
in the wide range of $x$ and $\phi$ for $s=-1$. {}From this figure, we 
see that we can acquire $r \approx 0.21$ for $n_\mathrm{s} \approx 0.96$. 
In this case, a typical value of $x$ is very large. 
For instance, for $N_e = 55$ and $n_\mathrm{s} = 0.9603$, 
we find $\phi \approx 34.7$, $x \approx 2.8 \times 10^9$, and 
eventually obtain $r \approx 0.212$. 

We explore the values of our initial parameters $\alpha$ and $C$. 
For clarity, we take $N_e = 55$. Using the definition of 
$x \equiv \alpha C$, we find 
\begin{equation} 
V_\mathrm{eff} \sim \frac{1.0 \times 10^{15}}{\alpha} \,.
\label{eq:IIB-2.27}
\end{equation}
This implies that the viable values are $C < 3.0 \times 10^{-6}$ and hence 
$\alpha > 1.0 \times 10^{15}$. 
We mention that for such a large value of $\alpha$, 
the term $1/\left(4\alpha\right)$ in the Friedman equation 
becomes small, and therefore it could play a role of the effective cosmological constant $\Lambda_\mathrm{eff}$. 
This term may lead to the late-time cosmic acceleration. 

\subsection{Equilibrium points}

Next, we investigate equilibrium points in the system. 
We have the system consisting of two dynamical equations (\ref{1.11}) and (\ref{1.12}) with a constraint equation (\ref{1.14}). 
To examine equilibrium points in this system, we need to rewrite the system of two second order differential equations as that of four first order differential equations. 
It is clear that this task is equivalent to explore the shape of the potential, namely, to find its extreme values and study their natures. 
We execute the numerical analysis by using the graphics of the potential for several values of parameters. 

\begin{center}
\begin{figure}[htbp]
  \includegraphics[width=8cm]{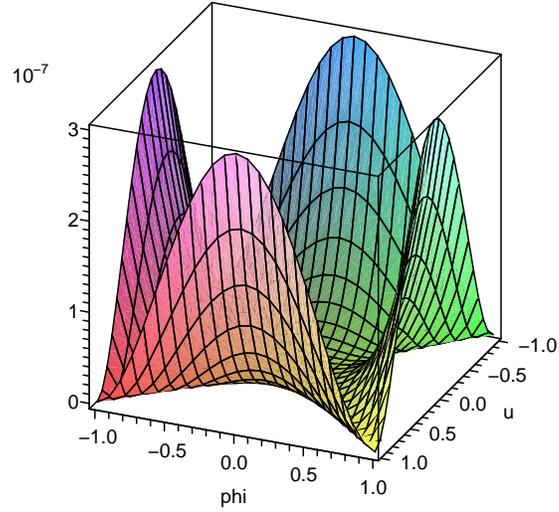}
 \caption{$V$ as a function of $\phi$ and $u$ in Eq.~(\ref{1.10}) near an extreme value for $\alpha=1.0 \times 10^{16}$, $C=3.0 \times 10^{-7}$, $s=-1$, and $J=C$.}
  \label{J=C}
\end{figure}
\end{center}

\begin{center}
\begin{figure}[htbp]
  \includegraphics[width=8cm]{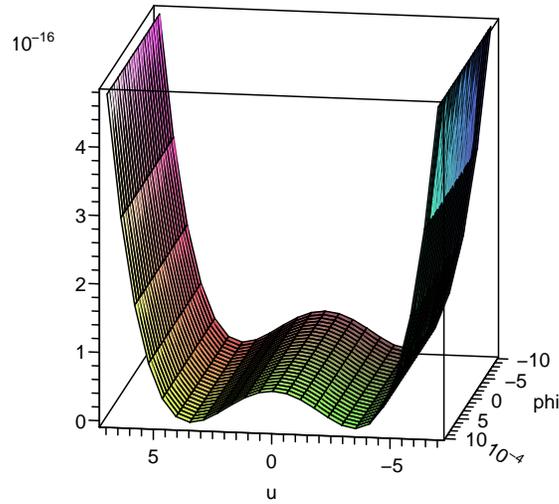}
 \caption{$V$ as a function of $\phi$ and $u$ in Eq.~(\ref{1.10}) around 
the minimum for $\alpha=1.0 \times 10^{16}$, $C=3.0 \times 10^{-7}$, $s=-1$, and $J=C(y^2-1)^{-2}$.}
  \label{J1}
\end{figure}
\end{center}

\begin{center}
\begin{figure}[htbp]
  \includegraphics[width=8cm]{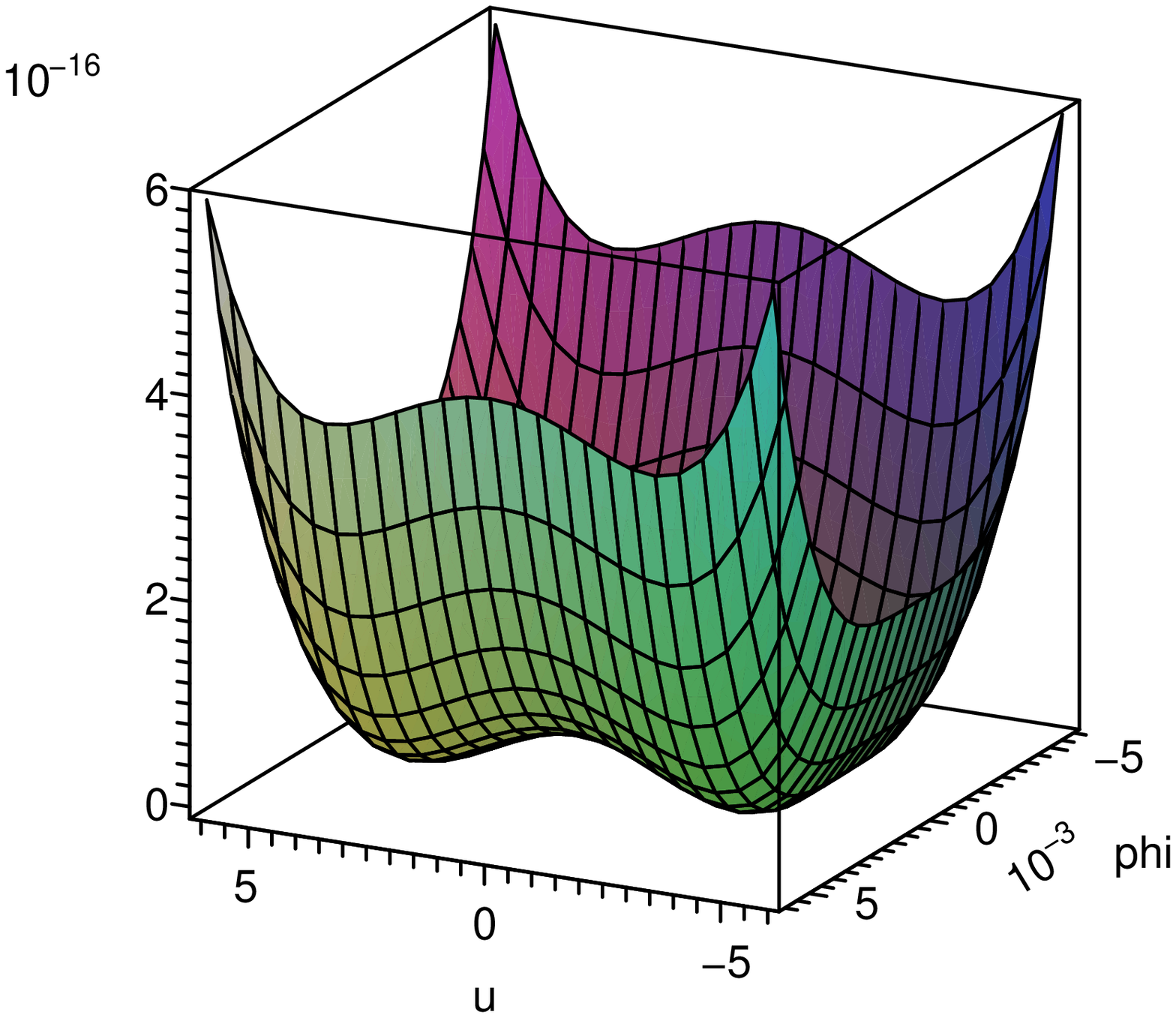}
 \caption{$V$ as a function of $\phi$ and $u$ in Eq.~(\ref{1.10}) around 
$\alpha=1.0 \times 10^{16}$, $C=3.0 \times 10^{-7}$, $s=-1$, and $J=C(y^2-1)^{-2}$.}
  \label{J2}
\end{figure}
\end{center}

For most of possible shapes of the function $J(y)$, 
there is no true minimum point. 
This is because the value of the potential in 
Eq.~(\ref{1.10}) on the lines $u^2=\phi^2$ is exactly 
equal to $1/\left(2\alpha\right)$. 
This fact is true for such kind of functions as $J=C$, $J=C(y^2-1)^m$ with $m$ a constant, $J=C\cos^2(y^2)$, $J=C\exp(y^2)$, and so on. 
The typical behavior of a potential for such kind of functions is drawn in Fig.~\ref{J=C}. Nevertheless, it is possible to construct functions $J$ for which the potential in Eq.~(\ref{1.10}) has a true minimum. 
For instance, we have 
\begin{equation}
J(y) = \frac{C}{\left(y^2-1\right)^2}\,. 
\label{1.24}
\end{equation}
The typical behaviors of the potential $V$ as a function of 
$\phi$ and $u$ in Eq.~(\ref{1.10}) 
are plotted in Figs.~\ref{J=C}--\ref{J2}. 

\subsection{Re-collapse and bounce solutions}

We explore the possible re-collapse and bounce solutions\footnote{Recently, cosmological scenarios to avoid the initial singularity in the early universe have proposed in various modified gravity theories~\cite{B-MG}.}. 
For such kinds of solutions, 
the relation $H=0$ has to be satisfied at some time. 
At the time, from Eqs.~(\ref{1.14}) and (\ref{1.15}), we have
\begin{equation}
\dot{H} =\frac{1}{2}V \,.
\label{1.24}
\end{equation}
It is clear that 
the condition $\dot H>0$ is necessary for bouncing solutions, 
whereas the condition $\dot H<0$ has to be met for re-collapsing solutions. 
Accordingly, it can be seen from the general form of the potential $V$ in Eq.~(\ref{1.10}) that for $J>0$, the possibility for 
bouncing solutions to exist becomes higher 
because the value of $V$ is defined to be positive. 
On the other hand, if the potential can have negative values, 
the existence of re-collapsing solutions is possible as well. 
However, in the case that the value of the potential is always positive, 
it is impossible for re-collapsing solutions to exist 
for, e.g., $J$ given by Eq.~(\ref{1.24}) with $\alpha>0$ and $C>0$.

Related to the bouncing solutions, we mention that 
the anti-gravity regime in the extended gravity theories
with the Weyl invariance~\cite{Bars:2011th, Bars:2011aa, B-BST-BST, Carrasco:2013hua, Bars:2013qna} including $F(R)$ gravity~\cite{Bamba:2014kza} 
has been examined.

\section{Inflationary cosmology}

In this section, we reconsider 
the theory whose action is described by Eq.~(\ref{1.1}) and 
build an inflationary model in another way. 

\subsection{Conformal transformation}

The action in Eq.~(\ref{1.1}) is written in the so-called Jordan frame. 
Instead of taking the gauge in Eq.~(\ref{1.3}), 
we first introduce an auxiliary field $\Phi$ and then 
make the conformal transformation, i.e., the Weyl re-scaling, 
of the metric from the Jordan frame to the Einstein frame~\cite{M-FM} 
\begin{equation}
g_{\mu\nu}=\Lambda \bar{g}_{\mu\nu}\,,
\label{3.3}
\end{equation}
where the bar shows the quantities in the Einstein frame. 
The Ricci scalar is transform as 
\begin{equation}
R=\Lambda^{-1}\left[ \bar{R} -3\Lambda^{-1} \bar{\Box}\Lambda +\frac{3}{2}\Lambda^{-2}(\bar{\nabla}\Lambda)^2  \right]\,. 
\label{3.4}
\end{equation}
By plugging this relation into Eq.~(\ref{1.2}) and setting 
\begin{equation}
\Lambda\left[\Phi+\frac{s}{12}\left(\phi^2-u^2\right)\right]=1\,,
\label{3.5}
\end{equation}
we find\footnote{Clearly, we see that $\sqrt{-g}=\Lambda^2\sqrt{-\bar{g}}$.}
\begin{eqnarray} 
S\Eqn{=}\int d^4x\sqrt{-\bar{g}}\left\{ \bar{R} -\frac{3}{2}\Lambda^{-2}(\bar{\nabla}\Lambda)^2
-\frac{1}{2\alpha} + \frac{s}{12\alpha}\Lambda(\phi^2-u^2)-\frac{s^2}{288\alpha}\Lambda^2(\phi^2-u^2)^2\right. 
\nonumber \\
&& 
\hspace{20mm}
\left.
{}+ \frac{s}{2}\Lambda\left[(\bar{\nabla}\phi)^2 - (\bar{\nabla} u)^2 \right] - \Lambda^2(\phi^2-u^2)^2J(y) \right\}\,,
\label{3.6}
\end{eqnarray}
where we have removed the auxiliary field $\Phi$ with 
the relation (\ref{3.5}). 
If we define $\Lambda$ as $\Lambda \equiv \e^{\lambda}$ with $\lambda$ 
a scalar field, Eq.~(\ref{3.6}) is represented as 
\begin{eqnarray}
S\Eqn{=}\int d^4x\sqrt{-\bar{g}}\left\{ \bar{R} -\frac{3}{2}(\bar{\nabla}\lambda)^2
-\frac{1}{2\alpha}\left[1-\frac{s}{12}\e^{\lambda}(\phi^2-u^2)\right]^2
\right. 
\nonumber \\
&& 
\hspace{20mm}
\left.
{}+ \frac{s}{2}\e^{\lambda}\left \{(\bar{\nabla}\phi)^2 - (\bar{\nabla} u)^2 \right \} - \e^{2\lambda}(\phi^2-u^2)^2J(y) \right\}\,.
\label{3.7}
\end{eqnarray}
For simplicity, from this point, we will not write the bar over the quantities and operators in the Einstein frame. 

We introduce the following form of the potential
\begin{equation}
V(\lambda,\phi,u,J)=\frac{1}{2\alpha}\left[1-\frac{s}{12}\e^{\lambda}(\phi^2-u^2)\right]^2 + \e^{2\lambda}(\phi^2-u^2)^2J(y)\,. 
\label{3.8}
\end{equation}
The equations of motion for the scalar fields $\lambda$, $\phi$, 
and $u$ are given by 
\begin{eqnarray}
&&
3\Box\lambda +\frac{s}{2}\e^{\lambda}[(\nabla\phi)^2-(\nabla u)^2] - V_{\lambda}=0\,,
\label{3.9} \\
&&
s\e^{\lambda}\Box\phi + s\e^{\lambda}\nabla^{\mu}\lambda\nabla_{\mu}\phi + V_{\phi}=0\,,
\label{3.10} \\ 
&&
s \e^{\lambda}\Box u 
+ s\e^{\lambda}\nabla^{\mu}\lambda\nabla_{\mu}u - V_{u}=0\,. 
\label{3.11}
\end{eqnarray}
Moreover, the gravitational field equation becomes 
\begin{equation}
R_{\mu\nu}-\frac{1}{2}Rg_{\mu\nu}+\frac{s}{2}\e^{\lambda}(\nabla_{\mu}\phi\nabla_{\nu}\phi-\nabla_{\mu} u \nabla_{\nu} u) -\frac{3}{2}\nabla_{\mu}\lambda\nabla_{\nu}\lambda 
+\frac{1}{2}g_{\mu\nu} \left \{ V - \frac{s}{2}\e^{\lambda}\left[ (\nabla\phi)^2-(\nabla u)^2\right ] +\frac{3}{2}(\nabla\lambda)^2
\right \} =0\,. 
\label{3.12}
\end{equation}
In the FLRW background, 
from Eqs.~(\ref{3.9})--(\ref{3.11}), we acquire 
\begin{eqnarray}
&&
3\ddot\lambda + 9H\dot\lambda +\frac{s}{2}\e^{\lambda}(\dot\phi^2-\dot u^2) + V_{\lambda}=0\,,
\label{3.15} \\
&&
s\e^{\lambda}\ddot\phi + 3s\e^{\lambda}H\dot\phi + s\e^{\lambda}\dot\lambda\dot\phi - V_{\phi}=0\,,
\label{3.16} \\
&&
s\e^{\lambda}\ddot u + 3s\e^{\lambda}H\dot u + s\e^{\lambda}\dot\lambda\dot u + V_{u}=0\,.
\label{3.17}
\end{eqnarray}
Furthermore, from Eq.~(\ref{3.12}), we get 
\begin{eqnarray}
&&
3H^2 + \frac{s}{4}\e^{\lambda}(\dot\phi^2-\dot u^2) -\frac{3}{4}\dot\lambda^2-\frac{1}{2}V=0\,,
\label{3.13} \\
&&
2\dot H + 3H^2 - \frac{s}{4}\e^{\lambda}(\dot\phi^2-\dot u^2) +\frac{3}{4}\dot\lambda^2-\frac{1}{2}V=0\,. 
\label{3.14}
\end{eqnarray}
%

\subsection{Inflationary model}

To calculate the observables for inflationary models including 
the spectral index of curvature perturbations $n_\mathrm{s}$ 
and the tensor-to-scalar ratio $r$, 
it is necessary for two scalar fields to be made constants. 
If the scalar field $\lambda$ is a constant and 
one of the scalar fields $\phi$ or $u$ plays a role of the inflaton field, 
we obtain the similar theory to that described in the previous section. 

We suppose that the scalar field 
$\lambda$ is the inflaton field and 
other two scalar fields are set to be constants as 
$\phi=\phi_0 (= \mathrm{constant})$ and $u=u_0 (= \mathrm{constant})$. 
It should be cautioned that both these scalar fields $\phi$ and $u$ 
cannot be made zero, 
because in this case, $V_\mathrm{eff}=1/(2\alpha)$ and 
hence the spectrum of the curvature perturbations is flat. 
Nevertheless, it is possible for one of these scalar fields (e.g., $u$) 
to be taken as zero, while the other has a non-zero value. 
In fact, however, 
it is impossible to construct any inflationary model with at least (even) one viable parameter in this way. 
This is a reason why the relations 
$\phi=\phi_0 \neq 0$ and $u=u_0 \neq 0$ with $u_0 \neq \pm \phi_0$ 
are forced to be set. 
Therefore, 
with Eqs.~(\ref{3.16}) and (\ref{3.17}), 
the conditions $V_{\phi}=0$, $V_u=0$ during inflation are expressed as 
\begin{eqnarray}
V_{\phi}\Eqn{=}-\frac{s}{6\alpha}\left[1-\frac{s}{12}\e^{\lambda}(\phi_0^2-u_0^2)\right]\e^{\lambda}\phi_0 + 4\e^{2\lambda}(\phi_0^2-u_0^2)\phi_0J(y_0) -\e^{2\lambda}(\phi_0^2-u_0^2)^2J'(y_0)\frac{u_0}{\phi_0^2}=0\,,
\label{3.18} \\ 
V_{u}\Eqn{=}\frac{s}{6\alpha}\left[1-\frac{s}{12}\e^{\lambda}(\phi_0^2-u_0^2)\right]\e^{\lambda}u_0 - 4\e^{2\lambda}(\phi_0^2-u_0^2)u_0J(y_0) +\e^{2\lambda}(\phi_0^2-u_0^2)^2J'(y_0)\frac{1}{\phi_0}=0\,.
\label{3.19}
\end{eqnarray}
There are the following cases in which these equations are realized:
\begin{itemize}
\item 
Case 1: \,\,\, 
The relation $K\equiv \left(s/12\right)\e^{\lambda}(\phi_0^2-u_0^2)\gg 1$ 
is met. In this case, the first terms in the brackets $[\,\,]$ 
in Eqs.~(\ref{3.18}) and (\ref{3.19}) 
may be neglected. Namely, 
the term with $\e^{\lambda}$ is much smaller 
than any other terms. Accordingly, all the other terms 
have the multiplier factor $\e^{2\lambda}$ incorporated in the same way, 
so that this overall factor can removed from the other terms. 

\item 
Case 2: \,\,\, 
The values of the first and second terms in the brackets $[\,\,]$ may be 
similar with each other, but the overall coefficient term 
$s/\left(6\alpha\right)$ is sufficiently small. As a consequence, 
the first terms in Eqs.~(\ref{3.18}) and (\ref{3.19}) are 
suppressed to be much smaller than all the other terms 
in these equations. 
\end{itemize}

Provided that (at least) one of Cases 1 and 2 is realized, 
namely, the first terms in Eqs.~(\ref{3.18}) and (\ref{3.19}) are 
negligible, 
by multiplying Eqs.~(\ref{3.18}) and (\ref{3.19}) by $u_0$ and $\phi_0$, 
respectively, and summing them, we acquire\footnote{Remind the facts that $\phi=\phi_0\neq 0$, $u=u_0\neq 0$, and $u_0 \neq \pm\phi_0$.}
\begin{equation}
J'(y_0)=0\,.
\label{3.20}
\end{equation}
Therefore, the function $J$ has to reach its extreme value when the argument becomes $y=y_0$. 
Moreover, by substituting this value into Eqs.~(\ref{3.18}) and (\ref{3.19}), 
we see that in Case 1, the relation $J(y_0)=-s^2/(288\alpha)$ has to be met, 
whereas in Case 2, the relation $J(y_0)=0$ has to be realized. 
Note also that 
a very wide class of functions may satisfy both these conditions. 
In the following, we study these two cases in more detail. 

\subsubsection{Case 1}

In Case 1, by combining the value of the function $J$ with the expression of the potential in Eq.~(\ref{3.8}), we have the effective inflaton potential
\begin{equation}
V_\mathrm{eff} (\lambda) =\frac{1}{2\alpha}\left[1-\frac{s}{6}\e^{\lambda}(\phi_0^2-u_0^2)\right]\,.
\label{3.21}
\end{equation}
This is some kind of the potential in the modified Higgs inflation model~\cite{Bezrukov:2007ep}. 
The number of e-folds in the slow-roll regime is expressed as 
\begin{equation}
N_{e}(\lambda)=\int^{\lambda_\mathrm{f}}_{\lambda}H(\lambda)\frac{d\lambda}{\dot\lambda}=\frac{3}{2}\int^{\lambda}_{\lambda_\mathrm{f}}\frac{V}{V_{\lambda}}d\lambda \,, 
\label{3.22}
\end{equation}
with $\lambda_\mathrm{f}$ the value of $\lambda$ at the end of inflation. 
If $\lambda \gg \lambda_\mathrm{f}$, 
the representation of $N_{e}(\lambda)$ 
for the effective potential in Eq.~(\ref{3.21}) becomes 
\begin{equation}
N_{e}(\lambda)=\frac{3}{2}\left[ \lambda +\frac{6}{s(\phi_0^2-u_0^2)\e^{\lambda}} \right]\,,
\label{3.23}
\end{equation}
or by taking into account the assumption that 
$K \equiv \left(s/12\right)\e^{\lambda}(\phi_0^2-u_0^2)\gg 1$, 
we may rewrite $N_{e}(\lambda)$ as 
\begin{equation}
N_{e}(\phi)\approx\frac{3}{2} \lambda \,.
\label{3.24}
\end{equation}

With Eqs.~(\ref{0.1}) and (\ref{0.2}), 
the slow-roll parameters can be described by using 
the effective potential as 
\begin{eqnarray}
\epsilon \Eqn{=} \frac{1}{3} 
\left( \frac{1}{V_\mathrm{eff}(\lambda)} \frac{dV_\mathrm{eff}(\lambda)}{d\lambda} \right)^2 \,, 
\label{3.25} \\
\eta \Eqn{=} \frac{2}{3} \left( \frac{1}{V_\mathrm{eff}(\lambda)} 
\frac{d^2 V_\mathrm{eff}(\lambda)}{d \lambda^2} \right)\,. 
\label{3.26}
\end{eqnarray}
Here, the coefficients of $1/3$ in Eq.~(\ref{3.25}) and $2/3$ in Eq.~(\ref{3.26}) originate from the non-canonical definition of 
the scalar field $\lambda$, namely, the coefficient of the kinetic term for 
$\lambda$ in the action in Eq.~(\ref{3.7}) is $3/2$, and not $1/2$. 
Eventually, $n_\mathrm{s}$ and $r$ are described as 
\begin{eqnarray}
n_\mathrm{s} \Eqn{=} \frac{\zeta^2 \e^{2\lambda}-60\zeta \e^{\lambda}+108}{3\left(\zeta \e^{\lambda}-6\right)^2}\,,
\label{3.28} \\
r \Eqn{=} \frac{16 \zeta^2 \e^{2\lambda}}{\left(\zeta \e^{\lambda}-6\right)^2}\,,
\label{3.27}   
\end{eqnarray}
where we have defined $\zeta$ as $\zeta \equiv s(\phi_0^2-u_0^2)$. 
Consequently, we see that $n_\mathrm{s}$ can be written as a function 
of $r$ as $n_\mathrm{s}=n_\mathrm{s}(r)$. 
However, the values of the parameter $\zeta$ 
lead to those of $n_\mathrm{s}$ and $r$, 
which are compatible with the observations. 
Indeed, for $N_e=60$ and $\zeta=0.01$, 
we obtain $r=0.004$ and $n_\mathrm{s}=0.9617$. 
In this case, we find $K=0.013$ ($\ll 1$). 
For $N_e=50$ and $\zeta=0.18$ 
(the value of $\zeta$ for $N_e=50$ 
becomes larger than that for $N_e = 60$ by 
one order of magnitude), 
we acquire $r=0.005$ and $n_\mathrm{s}=0.9568$. 
In this case, we have $K=0.015$ ($\ll 1$). 
As a result, we see that 
this case is inconsistent, because the initial assumption is $K \gg 1$, 
but the consequences 
suggest $K \ll 1$. It means that this case cannot be realized, 
in contrast with Case 2, as is shown next.

\subsubsection{Case 2}

In Case 2, as already mentioned above,  
we have the following relations for the function $J$: 
$J(y_0)=0$ and $J'(y_0)=0$. 
A simple example of such kind of the function $J$ is 
\begin{equation}
J(y)=C(y-y_0)^q \,, 
\quad 
q \geq 2 \,,
\label{3.29}
\end{equation}
where $q$ is a constant. 
The effective potential is given by 
\begin{equation}
V_\mathrm{eff} (\lambda) = \frac{1}{2\alpha}\left(1-\frac{\zeta}{12}
\e^{\lambda}\right)^2 \,,
\label{3.30}
\end{equation}
where we have used $\zeta \equiv s(\phi_0^2-u_0^2)$. 
In addition, the expression of $N_e$ reads 
\begin{equation}
N_{e}(\phi)=\frac{3}{4}\lambda +\frac{9}{\zeta \e^{\lambda}} \,. 
\label{3.31}
\end{equation}
Accordingly, we find 
\begin{eqnarray}
n_\mathrm{s} \Eqn{=} \frac{432 - 5 \zeta^2 \e^{2\lambda}-168\zeta \e^{\lambda}}{3(\zeta \e^{\lambda}-12)^2}\,,
\label{3.33} \\
r \Eqn{=} \frac{64 \zeta^2 \e^{2\lambda}}{3(\zeta \e^{\lambda}-12)^2} \,.
\label{3.32}  
\end{eqnarray}
Thus, it is seen that this theory also has only one parameter, and therefore 
we obtain $n_\mathrm{s}=n_\mathrm{s}(r)$. 
Through the numerical calculations, 
the value of $n_\mathrm{s}$ may be set near the value of $n_\mathrm{s}=0.96$ suggested by the observations. 
In fact, in the case that $N_e=60$, for 
$\zeta = 10^{-1}$, $10^{-2}$, $10^{-3}$, $10^{-4}$, $10^{-5}$, and $10^{-6}$, 
we have 
$n_\mathrm{s}=0.9652$, $0.9641$, $0.9629$, $0.9617$, $0.9604$, and 
$0.9589$, respectively. 
For all of these cases, the value of $r$ is about $r=0.004$. 
Moreover, similarly to the above results, 
in the case that $N_e=50$, for 
$\zeta = 10^{-1}$, $10^{-2}$, $10^{-3}$, $10^{-4}$, $10^{-5}$, and $10^{-6}$, 
we acquire $n_\mathrm{s}=0.9577$, $0.9561$, $0.9543$, $0.9524$, $0.9504$, and $0.9481$, respectively. 
For all of these cases, the value of $r$ is in the range of 
$r=0.005$--$0.008$. 
Consequently, in this case, 
we can obtain $n_\mathrm{s} \approx 0.96$ and $r < 0.11$. 
These results are compatible with the observations of the Planck 2015. 

Furthermore, we examine another aspect of this theory. 
If $N_e=60$ with $\zeta = 10^{-1}$, $10^{-2}$, $10^{-3}$, $10^{-4}$, $10^{-5}$, and $10^{-6}$, or if $N_e=50$ with these values of $\zeta$, 
namely, the same combinations of values of $N_e$ and $\zeta$ shown above, 
we obtain the estimation as 
$\left(\zeta/12\right) \e^{\lambda_\mathrm{f}} \thickapprox 0.14$--$0.17$. 
It means that for all the combinations of values of $N_e = 50$--$60$ (during inflation) and those of $\zeta$ described above, 
the value of the effective potential $V_\mathrm{eff}$ may be estimated as 
\begin{equation}
V_\mathrm{eff} \thickapprox\frac{0.7}{2\alpha} \,. 
\label{3.34}
\end{equation}
This relation implies that for sufficiently large values of $\alpha$, 
the values of the potential become under the Planck scale. 
This fact is in good agreement with our initial assumption 
that $s/\left(6\alpha\right) \ll 1$. 
As a result, it is considered that Case 2 is quite viable.

\section{Dynamical two scalar field model}

In this section, we explore the theory proposed in Sec.~II, 
whose action is given by Eq.~(\ref{1.1}), 
and consider how to realize inflation by using both dynamical two scalar fields $\phi$ and $u$\footnote{It should be noted that 
in the framework of this theory, infation is realized 
only for non-interacting scalar fields~\cite{Starobinsky:1986fxa}, 
whereas in the most general case, infation in this theory 
has not been realized yet, 
and it is not so clear to analyze the tensor-to-scalar ratio.}. 

%

The definitions of the slow-roll parameters in Eqs.~(\ref{0.1}) and 
(\ref{0.2}) are used to describe the slow-roll inflation 
in the present theory. 
In addition, the spectral index $n_\mathrm{s}$ of the curvature perturbations 
and the tensor-to-scalar ratio $r$ are supposed to be represented 
as Eqs.~(\ref{0.3}) and (\ref{0.4}), respectively. 
This assumption has been justified in Ref.~\cite{AW}. 

We consider the FLRW background. 
We start with the system of the equations (\ref{1.11}) and (\ref{1.12}), 
namely, (\ref{1.14}) and (\ref{1.15}) in the FLRW space-time. 
By imposing the standard slow-rolling conditions 
$\ddot u \ll \dot u H$, $\ddot\phi\ll\dot\phi H$, 
and $\dot\phi^2-\dot u^2\ll H^2$, 
the Friedmann equation and the equations of motion for $\phi$ and 
$u$ are reduced to be the following simple forms: 
\begin{eqnarray}
H^2 \Eqn{=} \frac{1}{6} V \,,
\label{4.1} \\ 
3sH \dot{\phi} \Eqn{=} V_{\phi} \,,
\label{4.2} \\
3sH \dot{u} \Eqn{=} -V_u \,.
\label{4.3}
\end{eqnarray}
First, we derive the number of $e$-folds by using the initial definition
\begin{equation}
N_{e}\equiv\int^{a_\mathrm{f}}_{a_\mathrm{i}}d\ln a= \int^{t_\mathrm{f}}_{t_\mathrm{i}}H dt \,,
\label{4.4}
\end{equation}
where $a_\mathrm{i}$ and $a_\mathrm{f}$ are the values of the scale factor $a(t)$ at the beginning $t_\mathrm{i}$ and end $t_\mathrm{f}$ of inflation, 
respectively. 
Moreover, we also have the following relation 
\begin{equation}
dV=V_udu+V_{\phi}d\phi=V_u\dot udt+ V_{\phi}\dot\phi dt \,. 
\label{4.5}
\end{equation}
By using Eqs.~(\ref{4.1})--(\ref{4.3}), we get 
\begin{equation}
N_{e}=\frac{s}{2}\int_{\phi,u}^{\phi_\mathrm{f}, u_\mathrm{f}}\frac{V \left(V_u du+V_{\phi}d\phi\right)}{V_{\phi}^2-V_u^2} \,, 
\label{4.6}
\end{equation}
with $\phi_\mathrm{f}$ and $u_\mathrm{f}$ the amplitudes of $\phi$ and $u$ 
at the end of inflation, respectively. 
It follows from the definition in Eq.~(\ref{0.1}) that 
$\epsilon$ is described as 
\begin{equation}
\epsilon =-\frac{\dot V}{2HV}= \frac{V_u^2 - V_{\phi}^2}{sV^2} \,.
\label{4.7}
\end{equation}
On the other hand, 
with the definition in Eq.~(\ref{0.2}) and Eq.~(\ref{4.1}), we find
\begin{equation}
\eta=-\frac{1}{4HV\dot V} \left(\dot V^2 + 2\ddot VV \right) 
=-\frac{2\left(V_{\phi}^2V_{\phi\phi}+V_u^2V_{uu}\right)}{sV\left( V_{\phi}^2-V_u^2 \right)} \,,
\label{4.8}
\end{equation}
where in deriving the last equality, we have used 
Eqs.~(\ref{4.2}) and (\ref{4.3}). 
The expressions (\ref{4.6})--(\ref{4.8}) can be used 
for any choice of the function $J$, but in the most general case, 
the calculations would be too cumbersome. 
In addition, 
the expression (\ref{4.6}) leads to only a relation 
between the scalar fields $\phi$ and $u$, and not  
both the values of $\phi$ and $u$ simultaneously.  
Hence, another additional assumption to determine both the values of 
$\phi$ and $u$ (e.g., $\phi \approx u$) is necessary, 
or one of the values of $\phi$ and $u$ may be regarded 
as a free parameter. 

In the simplest case $J=C$, it is possible to acquire the values of the main parameters analytically\footnote{Even in this simplest case, 
equations cannot be reduced to the one-field equations because of the kinetic terms.}. 
In this case, from Eq.~(\ref{4.6}), we find 
\begin{equation}
N_{e}=\frac{s}{16}(u^2-\phi^2)+\frac{3}{2}\ln(\phi^2-u^2)-\frac{432x}{288x+s^2}\ln\left[\left(288x+s^2\right)\left(\phi^2-u^2\right)-12s\right] \,. 
\label{4.10}
\end{equation}
Expressions for $\epsilon$ and $\eta$ takes the form
\begin{eqnarray}
\epsilon \Eqn{=} -\frac{16(\phi^2-u^2)}{s}\left\{ \frac{(288x+s^2)(\phi^2-u^2)-12s}{288x(\phi^2-u^2)^2+\left[12-s(\phi^2-u^2)\right]^2} \right\}^2 \,,
\label{4.11} \\
\eta \Eqn{=} - \left[ \frac{8}{s\left(\phi^2-u^2\right)} \right] \frac{\left(288x+s^2\right)\left(3\phi^4-2\phi^2u^2+3u^4\right)-12s\left(\phi^2-u^2\right)}{288x\left(\phi^2-u^2\right)^2+\left[12-s(\phi^2-u^2)\right]^2} \,.
\label{4.12}
\end{eqnarray}
With the expression of $N_e$ in (\ref{4.10}), 
the value of $(\phi^2-u^2)$ can be obtained. 
It determines the value of $\epsilon$, but 
that of $\eta$ cannot be found by using only that of $(\phi^2-u^2)$.  
Since $\epsilon (> 0)$ should be positive, from Eqs.~(\ref{4.10}) and 
(\ref{4.11}), we have $\phi^2-u^2>0$ and $s<0$. 
These relations and $x = \alpha C >0$ lead to the positive value of $\eta$. 

We estimate numerical values. 
For simplicity, we take $s=-1$. {}From Eqs.~(\ref{4.10}) and (\ref{4.11}), 
we get $(\phi^2-u^2)$ and $x$ for any interesting values of $N_e$ and 
$r=16\epsilon$. Thanks to the structure of the equations, we obtain 
the value of r as $r<0.11$.
On the other hand, it is seen that 
$n_\mathrm{s} = n_\mathrm{s} (r, \mathcal{G}^4)$, where we have 
defined a new parameter $\mathcal{G}^4 \equiv u^2\phi^2$. 
This relation yields 
the value of $n_\mathrm{s}$ consistent with the observations. 
Indeed, for $N_e=50$ and $x=10^{30}$, we have 
$\phi^2-u^2=2593$ and $r\simeq 0.1$. 
In this case, $n_\mathrm{s}$ is represented as 
\begin{equation}
n_\mathrm{s}=3.7 \times 10^{-9} \mathcal{G}^4+0.9815 \,. 
\label{4.13}
\end{equation}
Moreover, for $N_e=60$ and $x=10^{25}$, we acquire 
$\phi^2-u^2=2317$ and $r\simeq 0.11$. 
In this case, $n_\mathrm{s}$ is expressed as 
\begin{equation}
n_\mathrm{s} = 5.0 \times 10^{-9} \mathcal{G}^4 + 0.9793 \,.
\label{4.14}
\end{equation}
If the larger $x$ is, the smaller $r$ becomes, 
but the minimum value of 
$n_\mathrm{s}$ in both Eqs.~(\ref{4.13}) and (\ref{4.14}) increases. 
Equations (\ref{4.13}) and (\ref{4.14}) imply that 
it is impossible to for the pair of 
$n_\mathrm{s}$ and $r$ to take their values compatible with 
the observations. 
The consequence is the same for 
the different values of $s$ and $N_e$. 
Thus, for the simplest case of $J=C$,  
it is impossible to realize the Planck analysis 
with two dynamical scalar fields. 
However, even in this simplest case, 
the BICEP2 results may be realized. 
For $N_e=50$ and $x=10^6$, we obtain $\phi^2-u^2=1267$, $r\simeq 0.202$, 
and $n_\mathrm{s}=3 \times 10^{-8}\mathcal{G}^4+0.9621$. 
Therefore, the case of $\mathcal{G}^4 < 10^5$ is quite viable and consist with 
the BICEP2 experiment. 

As a result, in this simplest case (of function $J$), 
the Planck results cannot be realized. Nevertheless, 
the model with two dynamical scalar fields leads to 
significantly different results in comparison with 
the inflationary model with the single scalar field. 
There is another possibility to consider the more viable types of 
the function $J$, e.g., Eq.~(\ref{1.24}). 
In such a case, it is quite difficult to analyze the equations 
analytically 
(for example, $N_e$ is described as an integral equation).

\section{Graceful exit from inflation}

In this section, we investigate the graceful exit from inflation, 
namely, the instability of the de Sitter solution at the inflationary 
stage for the present theory. 
Especially, we demonstrate the instability of the de Sitter solution 
in the case of one dynamical scalar field in the Einstein frame. 
In this case, as shown in Sec.~III B 2, 
the spectral index and the tensor-to-scalar ratio 
can be compatible with the observations by the Planck satellite. 
We also examine the contribution of an $R^2$ term in the Jordan frame 
to the instability of the de Sitter solution during inflation. 
 
We set the perturbations of the Hubble parameter at the inflationary 
stage as 
\begin{equation}
H = H_\mathrm{inf} \left( 1 + \delta(t) \right) \,, 
\quad 
\left| \delta(t) \right| \ll 1
\label{eq:V.1} 
\end{equation}
where $H_\mathrm{inf} (>0)$ (= constant) 
is the Hubble parameter during inflation (whose value is positive), 
and $\delta(t)$ means the perturbations from the de Sitter solution. 
In the present case, only the scalar field $\lambda$ is dynamical, 
and $\phi$ and $u$ are constants. 
By taking the time derivative of the gravitational field equation (\ref{3.14}) 
with $V = V_\mathrm{eff}$, where $V_\mathrm{eff}$ is given by 
Eq.~(\ref{3.30}), we have 
\begin{equation}
2\ddot{H} +6H\dot{H} +\frac{3}{2} \dot{\lambda} \ddot{\lambda} + 
\frac{\zeta}{24\alpha} \left(1-\frac{\zeta}{12}
\e^{\lambda} \right)\e^{\lambda} \dot{\lambda} = 0 \,.
\label{eq:V.2} 
\end{equation}
Moreover, from Eq.~(\ref{3.15}) with $V = V_\mathrm{eff}$ 
(in Eq.~(\ref{3.30})), 
the equation of motion for $\lambda$ reads 
\begin{equation} 
3\ddot{\lambda} + 9H\dot{\lambda} - \frac{\zeta}{12\alpha} \left( 
1-\frac{\zeta}{12} \e^{\lambda} \right) \e^{\lambda} =0\,. 
\label{eq:V.3} 
\end{equation}
Since we consider the solution at 
the inflationary stage in the early universe, 
we take the limit $t \to 0$. 
In this limit, we could have an approximate solution $\lambda \approx \ln \left( H t \right)$ for Eq.~(\ref{eq:V.3}) with the quasi de Sitter solution 
$H \approx 1/\left(3t \right)$. 
By taking into account this solution, we see that 
in the limit $t \to 0$, the third term proportional to 
$\ddot{\lambda}$ 
in the left-hand side of Eq.~(\ref{eq:V.2})
is much larger than the fourth term, which is proportional only to 
$\dot{\lambda}$, and therefore the fourth term could be neglected. 

To examine the instability of the de Sitter solution, 
we represent $\delta(t)$ as 
\begin{equation} 
\delta(t) = \e^{\beta t} \,, 
\label{eq:V.4} 
\end{equation}
with $\beta$ a constant. 
When $\beta$ is positive, the de Sitter solution during inflation 
becomes unstable, and hence the universe can exit from 
the inflationary stage and then enter the reheating stage. 
This is because for $\beta > 0$, 
the amplitude of $\delta(t)$ increases in time. 
By substituting the expression of the perturbations in Eq.~(\ref{eq:V.1}) 
with Eq.~(\ref{eq:V.4}) 
into Eq.~(\ref{eq:V.2}) and using approximate solutions $\lambda \approx 
\ln \left( H t \right)$ and $H \approx 1/\left(3t \right)$, 
we find 
\begin{equation} 
2H_\mathrm{inf} \beta^2 + 6H_\mathrm{inf}^2 \beta + \frac{81}{2} H_\mathrm{inf}^3 = 0\,. 
\label{eq:V.5} 
\end{equation}
The solutions for this equation are obtained as 
\begin{equation} 
\beta_\pm = \frac{3\left(-1\pm\sqrt{10}\right)H_\mathrm{inf}}{2}\,. 
\label{eq:V.5} 
\end{equation}
As a consequence, 
we acquire the positive solution of $\beta = \beta_+ > 0$, 
and thus the universe can exit from the inflationary stage. 

We note that even if the other scalar field becomes dynamical, 
the procedure to examine the instability of the de Sitter solution 
is basically the same as the one demonstrated above. 
We examine the perturbations of the Hubble parameter by using 
the gravitational field equation with solutions for 
the equation of motions in terms of two dynamical scalar fields. 
Qualitatively, the form of the solution for the perturbations 
will be changed, but in principle there can exist 
a solution to represent the property that the de Sitter solution 
is unstable. 

The contribution of the $R^2$ term in the Jordan frame 
to the instability of the de Sitter solution is included 
in the scalar field $\lambda$ and its dynamics through the auxiliary 
field $\Phi$ in the Einstein frame. 
This fact can be seen from the action in Eq.~(\ref{1.2}), and 
the relation (\ref{3.5}) with $\Lambda = \e^{\lambda}$. 
Accordingly, it is considered that the $R^2$ term can be related to 
the graceful exit from inflation, i.e., the instability of 
the de Sitter solution to describe the slow-roll inflation.

\section{Conclusions}

In the present paper, we have studied inflationary cosmology 
in a theory where there exist two scalar fields non-minimally coupling to the Ricci scalar and an additional $R^2$ term. 
We have investigated the slow-roll inflation 
in the case of one dynamical scalar field and 
that of two dynamical scalar fields. 
We have analyzed the spectral index $n_\mathrm{s}$ of scalar mode of 
the density perturbations and the tensor-to-scalar ratio $r$ 
in comparison with the observations of the recent Planck and BICEP2 results. 

For the case of single dynamical scalar field in the Jordan frame, 
if the number of $e$-folds during inflation is $50 \leq N_e \leq 60$, 
we have $n_\mathrm{s} \approx 0.96$ and $r = \mathcal{O} (0.1)$. 
On the other hand, in the Einstein frame, 
for the case of one dynamical scalar field, 
we can acquire $n_\mathrm{s} \approx 0.96$ and $r < 0.11$. 
These are consistent with the Planck 2015 results. 
Furthermore, for the case of two dynamical scalar fields in 
the Jordan frame, when $50 \leq N_e \leq 60$, 
we obtain $n_\mathrm{s} \approx 0.96$ and $r = \mathcal{O} (0.1)$. 
As a result, we have found that in the present theory, 
the spectral index and the tensor-to-scalar ratio can be compatible 
with the recent Planck analysis. 
We have also shown that in the present theory, 
the de Sitter solution representing the inflationary stage 
is unstable, and therefore the universe can successfully exit from 
inflation. 
The $R^2$ term in the Jordan frame is considered to be 
related to the instability of the de Sitter solution, namely, 
the graceful exit from the inflationary stage. 

As the further developments on the cosmological consistent scenario in the present theory, it is possible to unify inflation in the early universe 
realized by the $R^2$ term and the late-time cosmic acceleration 
by the dark energy component of one of the scalar fields. 
In this theory, dark matter can also be explained by the other scalar field. 
To construct such a unified scenario is the significant purpose in this work. 
In addition, the consequences in multiple field inflation models 
seems to be more suitable for the observational data obtained form 
the Planck satellite than single field inflation models. 

It is remarked that 
the action in Eq.~(\ref{1.1}) may be generalized so that 
the conformal invariance can be broken by an arbitrary function of 
the $F(R)$ term. Thanks to the additional term in the gravity sector, 
the novel contributions to cosmology yield 
from the breaking of conformal invariance of the theory. 
Thus, we have more possibilities to realize 
the unified scenario of inflation, dark energy, and dark matter 
mentioned above. 

In fact, we may further extend the investigations in this work. 
We start from the theory with the conformally invariance and 
three scalar fields, and then add an $R^2$ term. 
We explicitly show the expressions of these actions below. 
In Ref.~\cite{Kallosh:2013daa}, a theory without an $R^2$ term and 
that with the general potential $(\phi^2 - u^2 - \psi^2)^2
J(u/\phi, u/\psi)$ have been explored. 
The action for the theory with the more specific potential $\psi^4 J(u/\phi, u/\psi)$ and without the $R^2$ term) is given by 
\begin{equation}
S=\int d^4x\sqrt{-g}\left\{  \frac{s}{2}\left[ \frac{(\phi^2-u^2 +\epsilon\psi^2)}{6}R + (\nabla\phi)^2 - (\nabla u)^2 +\epsilon(\nabla\psi)^2 \right]  - \psi^4J(\frac{u}{\phi},\frac{u}{\psi})\right\} \,, 
\label{5.1}
\end{equation}
where $\epsilon$ is a constant and $\psi$ is the additional third 
scalar field to two scalar fields $\phi$ and $u$ already introduced. 
If the $R^2$ term is added to Eq.~(\ref{5.1}), the representation of 
the action reads 
\begin{equation}
S=\int d^4x\sqrt{-g}\left\{ \frac{\alpha}{2} R^2 + \frac{s}{2}\left[ \frac{(\phi^2-u^2 +\epsilon\psi^2)}{6}R + (\nabla\phi)^2 - (\nabla u)^2 +\epsilon(\nabla\psi)^2 \right]  - \psi^4J(\frac{u}{\phi},\frac{u}{\psi})\right\}\,. 
\label{5.2}
\end{equation}
Moreover, when the general form of the potential is included and 
the signature of $\epsilon$ is opposite to that in the action 
in Eq.~(\ref{5.2}), the action is represented as 
\begin{equation}
S=\int d^4x\sqrt{-g}\left\{ \frac{\alpha}{2} R^2 + \frac{s}{2}\left[ \frac{(\phi^2-u^2 -\epsilon\psi^2)}{6}R + (\nabla\phi)^2 - (\nabla u)^2 -\epsilon(\nabla\psi)^2 \right]  - (\phi^2 - u^2 - \epsilon\psi^2)^2J(\frac{u}{\phi},\frac{u}{\psi})\right\}\,.
\label{5.3}
\end{equation}
The same analysis as in Secs.~II--IV 
may be adopted to the models in Eqs.~(\ref{5.1})--(\ref{5.3}) 
in which the additional third scalar field $\psi$ may play 
a role of the other specie of dark matter, 
or the additional contribution to the dark energy dominated stage or 
inflation.

\section*{Acknowledgments}

This work was partially supported by the JSPS Grant-in-Aid for 
Young Scientists (B) \# 25800136 (K.B.), 
MINECO (Spain) project FIS2010-15640 and FIS2013-44881 (S.D.O.), and 
the RFBR grant 14-02-00894A (P.V.T.).



\end{document}